\documentclass[a4paper,11pt]{article}

\usepackage{jheppub}
\usepackage[T1]{fontenc}
\usepackage{amsfonts}
\usepackage{amsmath}
\usepackage{amssymb,epsf}
\usepackage{latexsym}
\usepackage{graphicx,epsfig}
\usepackage{amssymb}

\title{\boldmath A chiral gauge-invariant model for Majorana neutrinos}

\author[a]{I. Alikhanov} 
\author[b]{and E.~A.~Paschos}

\affiliation[a]{Institute for Nuclear Research of the Russian Academy of Sciences,\\ 117312 Moscow, Russia}
\affiliation[b]{Department of Physics, TU Dortmund,\\ D-44221 Dortmund, Germany}

\emailAdd{ialspbu@gmail.com}
\emailAdd{paschos@physik.uni-dortmund.de}

\abstract{
The article investigates the possibility that right-handed neutrinos are Majorana particles embedded in an abelian multiplicative $U(1)_R$ factor, with the Lagrangian in the new factor being invariant under chiral gauge transformations. 
Majorana neutrinos couple to charged and neutral currents, producing new signatures to (anti)neutrino--electron elastic scattering and an additional term to coherent scattering of neutrinos on atomic nuclei. The model is ultraviolet complete and depends on one extra hypercharge correlating the reactions and providing upper bounds for the new coupling constant.
}


\begin{document}

\maketitle
\flushbottom

\section{Introduction}
The sector of the standard model that contains right-handed neutrinos (RHNs) appears, up to this time, to be incomplete. RHNs are necessary for introducing masses. Thus right-handed neutrinos are incorporated in Dirac and Majorana mass terms of the original Lagrangian and when gauge states are replaced by mass eigenstates they produce coupling of fermions to gauge bosons (charged currents). This happens in extensions of the standard model and one class of models embeds RHNs into a larger group~\cite{Fayet:1989mq,Appelquist:2002mw,Carena:2004xs,Langacker:2008yv,Fayet:2016nyc,DelleRose:2017xil,Correia:2019pnn,Correia:2019woz,Lindner:2018kjo}.  A typical prediction of the models is neutrinoless double beta decay~\cite{Bilenky:2014uka,Pas:2015eia,DellOro:2016tmg}.

In this article we extend the standard model with a multiplicative $U(1)_R$ group factor. The extended group is $SU(2)_L\times U(1)_Y\times U(1)_R$ with the Lagrangian for the $U(1)_R$ term being invariant under chiral gauge transformations. In the new group RHNs possess axial-vector charges and couple to the neutral gauge boson even though they are Majorana states. The new structure of the model is described in section~\ref{sec2}. In section~\ref{sec3} we demonstrate the cancellation of the appearing triangle anomalies.

Masses are generated by the standard Higgs doublet $\varphi(x)$ and a new singlet $\sigma(x)$. The doublet Higgs produces Dirac masses for quarks and leptons through Yukawa couplings, whose invariance determines the relations~\eqref{eq:2_23}--\eqref{eq:2_26} for the $U(1)_R$ charges~\cite{Fayet:1989mq,Appelquist:2002mw,Fayet:2016nyc}. Then considering the left-handed fermions to be singlets under $U(1)_R$ reduces them to one independent charge $Y_\varphi$. The doublet $\varphi(x)$ and singlet $\sigma(x)$ also contribute to the boson masses and generate a finite mass for the new gauge boson $X$. The symmetries are spontaneously broken preserving the symmetries of the currents, whose couplings to quarks and leptons are given in section~\ref{sec4}.

A basic requirement for an ultraviolet complete theory is the cancellation of triangle anomalies. The relations summarized in Table~\ref{tab_y} are necessary, but not sufficient, conditions for the theory to be ultraviolet finite.
In our case, the left-handed neutrinos do not couple to the $U(1)_R$ term and we select their charges to be zero: $Y_e=Y_{e_L}=0$ and $Y_u=Y_{u_L}=0$. Applying these conditions, a $\tau_{3R}$ symmetry survives for quarks and leptons, as shown in section~\ref{sec4}.  An explicit classification of the triangle diagrams demonstrates that the theory is ultraviolet finite.

Our interest in the extended theory was motivated by two questions: 

i. Are there models with Majorana neutrinos which couple to neutral gauge bosons?

ii. What is the structure of the new interactions and how they modify neutrino induced reactions?

 In section~\ref{sec5} we derive constraints introduced by the $\rho$ parameter and obtain a bound for the new gauge coupling. In section~\ref{sec6} we discuss the existence of a light $X$ boson and compute the changes introduced into antineutrino--electron elastic scattering. In addition, Majorana states may be produced in beam-dump experiments and in spallation sources inducing elastic scattering
 \begin{equation}
N_M+e^-\rightarrow N_M +e^-\label{eq:1in}
\end{equation}
with $N_M$ being a Majorana neutrino, and the reaction is mediated by the exchange of an $X$ boson. An analysis for the differential cross section restricts the coupling constants as a function of the $X$ boson mass.  

In section~\ref{sec7}, we consider the coherent scattering of the new particles on atomic nuclei and determine the range of parameters. The new interactions are mediated either by the $Z_\mu$ boson acquiring small components through mixing or by the presence of Majorana neutrinos in the beams. The latter requires the production of new fluxes in beam dump experiments or in cosmic rays. The reactions in sections~\ref{sec6} and~\ref{sec7} depend on the same $U(1)_R$ charge and their cross sections are related. Section~\ref{conclusion} contains the summary and conclusions.

\section{Description of the model\label{sec2}}

In the article, all states of the standard model retain their couplings and assignments in the group $SU(2)_L\times U(1)_Y$. Only right-handed neutrinos couple to a new gauge boson in $U(1)_R$ as follows

\begin{equation}
\mathcal{L}_n=\bar\Psi_Ri\gamma^\mu\left(\partial_\mu+ig_X\gamma^5\frac{Y_\Psi}{2}\hat X_\mu\right) \Psi_R+h.c.\label{eq:2_1}
\end{equation}
We require $\mathcal{L}_n$ to be invariant under $\gamma^5$-gauge (chiral) transformations

\begin{equation}
\Psi_R'(x)=e^{i\gamma^5H(x)}\Psi_R(x) \label{eq:2_3}
\end{equation}
and
\begin{equation}
\bar\Psi_R'(x)=\bar\Psi_R(x) e^{i\gamma^5H(x)}, \label{eq:2_4}
\end{equation}
where $H(x)$ is an $x$-dependent function. 
The Lagrangian 

\begin{equation}
\mathcal{L'}_n=\bar\Psi_R'(x) i\gamma^\mu\left(\partial_\mu+ig_X\gamma^5\frac{Y_\Psi}{2}\hat X'_\mu\right) \Psi_R'(x)+h.c.\label{eq:2_2}
\end{equation}
remains invariant when $\hat X'_\mu$ satisfies 

\begin{equation}
ig_X\gamma^5\frac{Y_\Psi}{2}\hat X_\mu=e^{-i\gamma^5H}\left(i\gamma^5\partial_\mu H+ig_X\gamma^5\frac{Y_\Psi}{2}\hat X'_\mu\right)e^{i\gamma^5H}. \label{eq:2_5}
\end{equation}
Thus the field transformation is

\begin{equation}
\delta\hat X_\mu=\hat X'_\mu-\hat X_\mu=-\frac{\partial_\mu H}{g_X}. \label{eq:2_7}
\end{equation}

For the non-abelian case the $\gamma^5$-gauge transformations are defined as 

\begin{equation}
\Psi'_R(x)=e^{i\gamma^5\vec\lambda\cdot\vec H}\Psi_R(x)\,\,\,\,\text{and}\,\,\,\,\bar\Psi'_R(x)=\bar\Psi'_R(x)e^{i\gamma^5\vec\lambda\cdot\vec H} \label{eq:2_8pr}
\end{equation}
with $\lambda_i$ being the generators of the group, $H^i=H^i(x)$ are functions of $x$. Equation~\eqref{eq:2_7} is now replaced by

\begin{equation}
\gamma^5g_X\delta\hat X^\alpha_\mu=-\gamma^5\partial_\mu H^\alpha(x)-g_X[\lambda^i,\lambda^\beta]H^i(x)\hat X^\beta_\mu. \label{eq:2_7repl}
\end{equation}
The first term on the right-hand side has a $\gamma^5$ but the second does not. 
Thus a $\gamma^5$-gauge transformation can be carried out only for an abelian theory, when the second term vanishes. We add the new term $\mathcal{L}_n$ from~\eqref{eq:2_1} to the fermion sector of the theory.

\subsection{The gauge sector}

The new field $X_\mu$ couples to the standard Higgs doublet $\varphi(x)$ through the covariant derivative

\begin{equation}
D_{\mu}\varphi(x)=\left(\partial_\mu+ig\frac{\tau_3}{2}\hat W^{3}_\mu+ig'\frac{Y}{2}\hat B_{\mu}+ig_X\frac{Y_\varphi}{2}\hat X_{\mu}\right)\varphi(x). \label{eq:2_8}
\end{equation}
Mass terms for the three bosons are generated when $\varphi(x)$ acquires a vacuum expectation value. An additional scalar Higgs $\sigma(x)$ is also introduced to provide another mass term $(1/4)g_X^2Y'^2v_0^2$ for the $X_\mu$ field with $Y'$ being the charge of $\sigma(x)$ in the $U(1)_R$ group. This brings a non-zero mass for the new gauge boson. After symmetry breaking the square of the mass matrix for the fields $\hat W^3_\mu$, $\hat B_\mu$ and $\hat X_\mu$ has the form~\cite{Fayet:1989mq,Appelquist:2002mw,Fayet:2016nyc,Correia:2019pnn,Lindner:2018kjo}

\begin{equation}
M^2=\frac{1}{4}\left(\begin{array}{ccc}
g^2 & -gg' & -gg_XY_\varphi \\
-gg' & g'^2 & g'g_XY_\varphi\\
-gg_XY_\varphi & g'g_XY_\varphi & g_X^2{\tilde{Q}_X}^2\\
\end{array}
\right)\frac{v^2}{2},
\label{eq:2_9}
\end{equation}
where $Y_X=Y_\varphi$ is the hypercharge of the Higgs doublet in $U(1)_R$ and
\begin{equation}
M^2_{33}=\frac{1}{4}g_X^2{\tilde{Q}_X}^2\frac{v^2}{2}=\frac{1}{4}\left(g_X^2Y_\varphi^2+g_X^2Y'^2\frac{v_0^2}{v^2}\right)\frac{v^2}{2}.\label{eq:2_10}
\end{equation}
In this specific extension the photon remains massless and electromagnetism is not affected.

The diagonalization of the mass matrix is well known, with the electromagnetic field retaining its form in the standard model,

\begin{equation}
A_{\mu}(x)=s_W\hat W^{3}_\mu(x)+c_W\hat B_{\mu}(x).\label{eq:2_11}
\end{equation}
For the computation of the neutral current, it is useful to express the gauge fields in terms of the physical fields $A_\mu$, $Z_\mu$ and $X_\mu$:

\begin{equation}
\hat W^3_\mu=s_WA_{\mu}+c_\alpha c_WZ_\mu+s_\alpha c_WX_\mu,\label{eq:2_12}
\end{equation}

\begin{equation}
\hat B_\mu=c_WA_{\mu}-c_\alpha s_WZ_\mu-s_\alpha s_WX_\mu,\label{eq:2_13}
\end{equation}

\begin{equation}
\hat X_\mu=-s_\alpha Z_{\mu}+c_\alpha X_\mu,\label{eq:2_14}
\end{equation}
where $s_W=\sin\theta_W$, $c_W=\cos\theta_W$ with $\theta_W$ being the Weinberg angle and  $s_\alpha=\sin\alpha$, $c_\alpha=\cos\alpha$ define a new mixing angle $\alpha$ given by~\cite{Fayet:1989mq}

\begin{equation}
c_\alpha^2=\frac{g^2+g'^2}{g^2+g'^2+g_X^2Q_X^2}\approx\frac{g^2+g'^2}{g^2+g'^2+g_X^2Y_\varphi^2}.\label{eq:2_15}
\end{equation}
The diagonalization of the mass matrix~\eqref{eq:2_9} -- in the limit $(v_0^2/v^2)\ll1$ -- provides the relation

\begin{equation}
g_Xc_\alpha Y_\varphi=\frac{gs_\alpha}{c_W}\label{eq:2_16}
\end{equation}
which will be useful later on.

\subsection{Fermion sector}

Lepton and quark masses are generated from their Yukawa couplings with standard Higgs doublet and in addition a new $SU(2)_L$-singlet $\sigma(x)$, when they acquire vacuum expectation values. The interaction for the three generations of leptons are

\begin{equation}
\mathcal{L}_Y=-Y^{ij}_e\bar l_{Li}\varphi e_{Rj}-Y^{ij}_\nu\bar l_{Li}\tilde{\varphi}\nu_{Rj}-Y^{ij}_\sigma\bar\nu_{Ri}\nu^C_{Rj}\sigma(x)+h.c.\label{eq:2_17}
\end{equation}
Here  $Y_e$, $Y_\nu$ and $Y_\sigma$ are $3\times3$ matrices with a summation over repeated indices. When the VEVs appear they produce a fermion mass matrix  with Dirac and Majorana terms. As stated previously we discuss the mass for the first generation. The explicit terms in~\eqref{eq:2_17} together with their charged-conjugate parts produce a type-I mass matrix with seesaw structure:

\begin{equation}
m_F=\left(\bar\nu_L \,\,\,\overline{ \nu_R^C}\right)\left(
\begin{array}{cc}
0&m_D\\
m_D^T&M\\
\end{array}
\right)\left(
\begin{array}{c}
\nu^C_L\\
\nu_R\\
\end{array}
\right)+h.c.\label{eq:2_18}
\end{equation}
The two mass eigenstates are

\begin{equation}
\Psi_1(x)=\nu_L(x)-\frac{m_D}{M}\frac{1}{\sqrt{2}}\left(\nu_R+\nu^C_R\right),\label{eq:2_19}
\end{equation}

\begin{equation}
\Psi_2(x)=\frac{m_D}{M}\nu_L(x)+\frac{1}{\sqrt{2}}\left(\nu_R+\nu^C_R\right)\label{eq:2_20}
\end{equation}
with $\Psi_2(x)$ being the heavier state, which has a large component of right-handed neutrinos and a smaller one with left-handed neutrinos. Inverting these equations one obtains

\begin{equation}
\nu_L(x)=\Psi_1(x)+\frac{m_D}{M}\Psi_2(x),\label{eq:2_21}
\end{equation}

\begin{equation}
\frac{1}{\sqrt{2}}\left(\nu_R+\nu^C_R\right)=-\frac{m_D}{M}\Psi_1(x)+\Psi_2(x).\label{eq:2_22}
\end{equation}
Substituting $\nu_L$ into the charged current couplings produces the interaction for neutrinoless double beta decays. The second equation~\eqref{eq:2_22} will be useful for computing neutral current interactions.

The invariance of the Dirac mass terms in~\eqref{eq:2_17} under $U(1)_R$ transformations restricts the charges of left- and right-handed leptons and quarks. The sum of charges in each trilinear must be zero, thus giving the relations~\cite{Fayet:2016nyc}

\begin{equation}
Y_{u_R}=Y_{u_L}+Y_\varphi, \label{eq:2_23}
\end{equation}

\begin{equation}
Y_{d_R}=Y_{u_L}-Y_\varphi, \label{eq:2_24}
\end{equation}

\begin{equation}
Y_{e_R}=Y_{e_L}-Y_\varphi. \label{eq:2_25}
\end{equation}
In our case a Dirac mass for the neutrinos is generated from the Yukawa coupling to the right-handed neutrino, giving

\begin{equation}
Y_{N_R}=Y_{e_L}+Y_\varphi. \label{eq:2_26}
\end{equation}
In this manner there are three independent $U(1)_R$ fermionic charges for each generation summarized in Table~\ref{tab_y}.

\begin {table}
\begin{center}
\begin{tabular}{ |c|c|c| }
  \hline
& $Y_{SM}$ & $U(1)_R$ \\\hline
   $\nu_L$, $e_L$ & -1 & $Y_e$ \\
	    $u_L$, $d_L$ & 1/3 & $Y_u$ \\
			$N_M$ & 0 & $Y_{N_R}=Y_e+Y_\varphi$ \\
     $e_R$ & -2 & $Y_{e_R}=Y_e-Y_\varphi$ \\
     $u_R$ & 4/3 & $Y_{u_R}=Y_u+Y_\varphi$ \\
     $d_R$ & -2/3 & $Y_{d_R}=Y_u-Y_\varphi$ \\
		 \hline
		 $\varphi$ & 1 & $Y_\varphi$ \\ 
  \hline
\end{tabular}
\caption {$U(1)_R$ charges for leptons and quarks.}\label{tab_y}
\end{center}
\end {table}
In the table we also included the standard model hypercharges denoted by $Y_{SM}=2(Q-\tau_3/2).$ We mention that whenever all charges are active, the charge assignments in Table~\ref{tab_y}, together with the relation $Y_e+3Y_u=0$ are sufficient for the cancellation of anomalies. Note, also, that Majorana mass terms of the form $M\overline{\nu^C_R}\nu_R$ violate the $U(1)_R$ symmetry. However in~\eqref{eq:2_17} the $\overline{\nu^C_R}\nu_R$ term couples to the scalar $\sigma(x)$ so that the phase generated by the transformation is absorbed into the phase of the $\sigma(x)$ field.

The fact that several $U(1)_R$ charges are arbitrary permits a two parameter family of $U(1)_R$ models. This was already noticed and discussed briefly in articles~\cite{Appelquist:2002mw,Correia:2019pnn} . A simple example identifies the charges of the leptons with a new $\tau_{3R}$ symmetry which implies $Y_{e_R}=Y_{d_R}=-Y_{\varphi}$~\cite{Appelquist:2002mw}. This will be our choice with which we discuss the cancellation of anomalies. Another choice identifies $U(1)_R$ with $U(1)_{B-L}$~\cite{Fayet:1989mq,Appelquist:2002mw,Carena:2004xs,Langacker:2008yv} which implies $Y_\varphi=0$; in this case $X_\mu$ does not mix with the standard model bosons.

\section{Cancellation of anomalies\label{sec3}}

In addition to the conditions required for the generation of fermion masses, there are restrictions on the hypercharges from the cancellation of triangle anomalies. In the $U(1)_R$ extension of the standard model there are general conditions~\cite{Appelquist:2002mw} which are consistent with the conditions in Table~\ref{tab_y}. In our case the left-handed states of quarks and leptons do not couple to the $U(1)_R$ bosons and we must set $Y_e=Y_u=0$. This reduces the three independent parameters of the table to one

\begin{equation}
Y_{e_R}=Y_{d_R}=-Y_{\varphi}\,\,\,\text{and}\,\,\,Y_{N_R}=Y_{u_R}=Y_{\varphi}.\label{eq:3_1}
\end{equation}
In addition the Majorana couplings are "axial-vector" and in the loops when the momenta of integrations are much larger than the masses, their contributions are equivalent to those for Dirac fermions with a $\gamma^5\gamma^\mu$ vertex.

We classify the diagrams according to the number of external $X_\mu$ fields.

\begin{figure}
\centering
\includegraphics[width=1.1\textwidth]{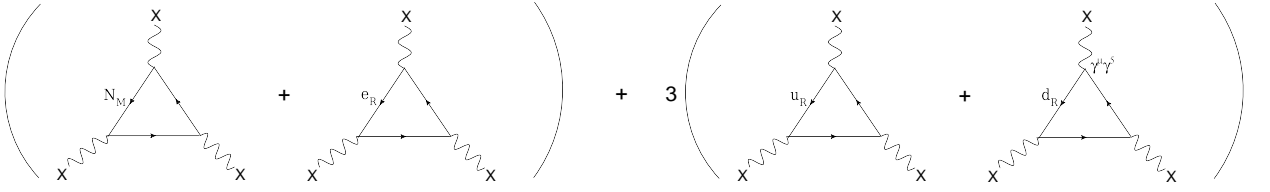}
\caption{Triangle diagrams with three $X_\mu$ external fields.}
\label{fig:1}
\end{figure}

1) Diagrams with three external fields are shown in Fig.~\ref{fig:1}. They cancel by virtue of the condition

\begin{equation}
Y_{N_R}^3+Y_{e_R}^3=(-Y_{\varphi})^3+Y_\varphi^3=0\,\,\,\,\text{and}\,\,\,\,Y_{u_R}^3+Y_{d_R}^3=0.\label{eq:3_2}
\end{equation}

\begin{figure}
\centering
\includegraphics[width=0.8\textwidth]{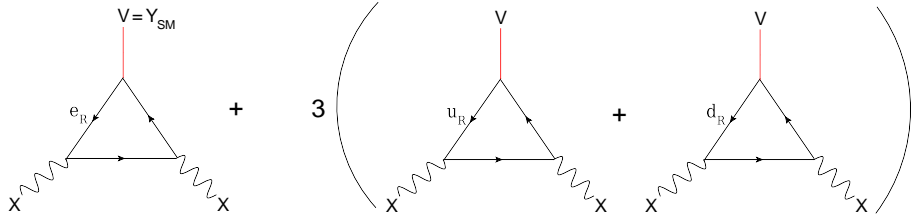}
\caption{Triangle diagrams for the case $Y_{SM}X^2$.}
\label{fig:2}
\end{figure}

2) For the case $Y_{SM}X^2$, the cancellation of anomalies requires the condition

\begin{equation}
V_{e_R}Y_{e_R}^2+3(V_{u_R}Y_{u_R}^2+V_{d_R}Y_{d_R}^2)=-2Y_{e_R}^2+3\left(\frac{4}{3}Y_{u_R}^2-\frac{2}{3}Y_{d_R}^2\right)=0\label{eq:3_3}
\end{equation}

as it follows from Fig.~\ref{fig:2}.

\begin{figure}
\centering
\includegraphics[width=0.8\textwidth]{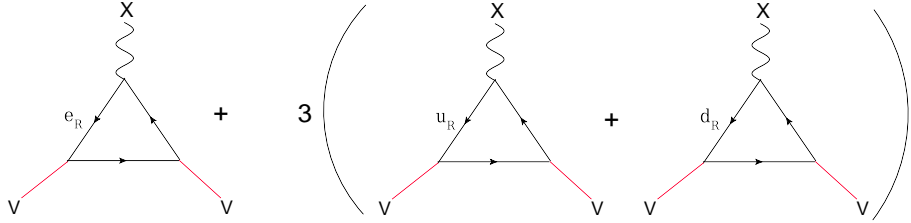}
\caption{Triangle diagrams for the case $Y_{SM}^2X$.}
\label{fig:3}
\end{figure}

3) For $Y_{SM}^2X$, the conditions required for the cancellation are

\begin{equation}
V_{e_R}^2Y_{e_R}+3(V_{u_R}^2Y_{u_R}+V_{d_R}^2Y_{d_R})=4Y_{e_R}+\left(\frac{16}{3}Y_{u_R}+\frac{4}{3}Y_{d_R}\right)=0.\label{eq:3_4}
\end{equation}
The corresponding diagrams are shown in Fig.~\ref{fig:3}.

4) For $(QCD)^2X$, the cancellations require

\begin{equation}
Y_{u_R}+Y_{d_R}=0.\label{eq:3_5}
\end{equation}

\begin{figure}
\centering
\includegraphics[width=1.1\textwidth]{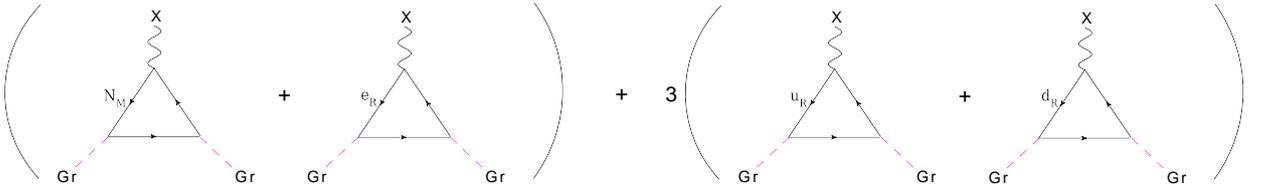}
\caption{Triangle diagrams for the case $(Gravity)^2X$.}
\label{fig:4}
\end{figure}

5) For $(Gravity)^2X$, the diagrams in Fig.~\ref{fig:4} lead to the condition

\begin{equation}
Y_{N_R}+Y_{e_R}+3\left(Y_{u_R}+Y_{d_R}\right)=0.\label{eq:3_6}
\end{equation}

As we mentioned earlier the charge assignments in Table~\ref{tab_y} with the condition $Y_e+3Y_u=0$ are sufficient for the cancellation of anomalies. Our approach satisfies this requirement with additional condition $Y_e=Y_u=0$.

\section{Structure of neutral currents\label{sec4}}

The neutral current interactions for left-handed and right-handed fermion fields are prescribed by the Lagrangian

\begin{equation}
\mathcal{L}_{NC}=\bar\Psi_Li\gamma^\mu\left[ig\frac{\tau_3}{2}\hat W^{3}_\mu+ig'\frac{Y}{2}\hat B_{\mu}\right]\Psi_L+\bar\Psi_Ri\gamma^\mu\left[ig'\frac{Y}{2}\hat B_{\mu}+ig_X\frac{Y_\Psi}{2}\gamma^5\hat X_{\mu}\right]\Psi_R. \label{eq:4_1}
\end{equation}
The hat on the fields indicates flavor fields which are replaced, with the help of~\eqref{eq:2_12}--\eqref{eq:2_14}, with physical gauge fields. The complete neutral current interaction is written  in matrix form:

\begin{equation}
\left(
\begin{array}{c}
\mathcal{L}_{NC}(Z)\\
\mathcal{L}_{NC}(X)\\
\end{array}
\right)=-\left(
\begin{array}{cc}
Z_\mu\frac{gc_\alpha}{c_W} &\,\, -Z_\mu g_Xs_\alpha\\
X_\mu\frac{gs_\alpha}{c_W} &\,\, X_\mu g_Xc_\alpha\\
\end{array}
\right)\left(
\begin{array}{c}
\bar\Psi_L\gamma^\mu\frac{\tau_3}{2}\Psi_L-s_W^2\sum\limits_\Psi \bar\Psi\gamma^\mu Q\Psi \\
\sum\limits_{\Psi_R} \bar\Psi_R\gamma^\mu\gamma^5\frac{Y_\Psi}{2}\Psi_R\\
\end{array}
\right).\label{eq:4_2}
\end{equation}
The upper component in the column matrix on the right-hand side has the structure of the standard model and the lower component is the new contribution from the initial $\hat X_\mu$ field. The $2\times2$ matrix emphasizes the mixing between the gauge bosons~\cite{DelleRose:2017xil}. For the first generation of quarks, the left-handed state is $\Psi_L^T=(u_L\,\,d_L)$ and the right-handed states are $u_R$ and $d_R$. The summation $\sum\limits_\Psi$ is over the charged states and $\sum\limits_{\Psi_R}$ runs over $u_R$ and $d_R$ carrying the $U(1)_R$ charges $Y_\varphi$ and $-Y_\varphi$, respectively. Introducing them in~\eqref{eq:4_2} we obtain the $\mathcal{L}_{NC}(X)$ interaction for quarks

\begin{eqnarray}
-\mathcal{L}_{NC}(X)=X_\mu\frac{g_Xc_\alpha}{4}Y_\varphi\left[\bar u\gamma^{\mu}(1-\gamma^5)u-\bar d\gamma^{\mu}(1-\gamma^5)d-s_W^2\left(\frac{8}{3}\bar u\gamma^\mu u-\frac{4}{3}\bar d\gamma^\mu d\right)\right]\nonumber\\+X_\mu\frac{g_Xc_\alpha}{4}Y_\varphi\left[\bar u\gamma^{\mu}(1+\gamma^5)u-\bar d\gamma^{\mu}(1+\gamma^5)d\right]\nonumber\\
=X_\mu\frac{g_Xc_\alpha}{4}Y_\varphi\left[2(\bar u\gamma^{\mu}u-\bar d\gamma^{\mu}d)-s_W^2\left(\frac{8}{3}\bar u\gamma^\mu u-\frac{2}{3}\bar d\gamma^\mu d\right)\right].\label{eq:4_3}
\end{eqnarray}
The final couplings are vector in ordinary space and their isospin content is mostly isovector.

For the leptons we use left-handed states and for the right-handed states we keep the leading term for Majorana neutrinos obtaining the couplings to $X_\mu$:

\begin{eqnarray}
-\mathcal{L}^{\text{leptons}}_{NC}(X)=X_\mu\frac{g_Xc_\alpha}{4}Y_\varphi\left[\bar u_\nu\gamma^{\mu}(1-\gamma^5)u_\nu-\bar u_e\gamma^{\mu}(1-\gamma^5)u_e+4s_W^2\bar u_e\gamma^\mu u_e\right.\nonumber\\\left.+(\bar u_N\gamma^\mu\gamma^5u_N-\bar u_e\gamma^\mu(1+\gamma^5)u_e)\right]\nonumber\\
=X_\mu\frac{g_Xc_\alpha}{4}Y_\varphi\left[\bar u_N\gamma^{\mu}\gamma^5u_N+\bar u_\nu\gamma^{\mu}(1-\gamma^5)u_\nu-2(1-2s_W^2)\bar u_e\gamma^{\mu}u_e\right].\label{eq:4_4}
\end{eqnarray}

There are two "axial-vector" terms, one for the Majorana particle and the other for the normal neutrino. When both of them are present in a beam they contribute incoherently. The term $\bar u_\nu\gamma^{\mu}(1-\gamma^5)u_\nu$ comes from the traditional neutrino and was generated from the mixing of $X_\mu$ with the $Z_\mu$ boson.

\section{Constrains on the mixing angle and the new couplings\label{sec5}}
The model is subject to constrains. The eigenvalues of the mass matrix~\eqref{eq:2_10} are complicated algebraic functions, but the leading terms are simple and sufficient for the analysis. The $Z$ boson mass has a new contribution 

\begin{equation}
m_Z=\frac{1}{2}v\sqrt{g^2+g'^2+g_X^2Y_\varphi^2}.
\end{equation}
A limit on the mixing angle is obtained from the $\rho$ parameter which is modified

\begin{equation}
\frac{m_W^2}{m_Z^2}=\frac{g^2}{g^2+g'^2+g_X^2Y_\varphi^2}=c_W^2c_\alpha^2.\label{eq:23n}
\end{equation}
The experimental value for the ratio 
\begin{equation}
\rho=\frac{m_W^2}{m_Z^2c_W^2}=c_\alpha^2=1.00037\pm0.00023
\end{equation}
restricts the mixing angle. To within two standard deviations\footnote{The value of $c_\alpha^2$ is very close to unity so that the strengths of the neutral relative to the charged current coupling are for practical purposes the same.}
\begin{equation}
0.99991\leq c_\alpha^2\leq1\label{eq:5_4}
\end{equation}
restricting $s_\alpha^2<8\times10^{-5}$ or $s_\alpha\leq9\times10^{-3}\approx0.01$~\cite{Correia:2019woz}. 

Combining~\eqref{eq:5_4} with the relation~\eqref{eq:2_16} leads to the upper bound
\begin{equation}
g_Xc_\alpha Y_\varphi<2.3\times10^{-3}.\label{eq:restr_c}
\end{equation}
We show later on that the upper bound produces substantial effects on antineutrino--electron scattering and must be restricted even more.

The coupling of $Z_\mu$ to $\bar N_MN_M$ brings a new decay. The currently available experimental data cannot eliminate a partial decay width of 0.5~MeV. These are decays to weakly interacting particles like $N_M$, which give the bound

\begin{equation}
g_Xs_\alpha Y_\varphi<0.06.\label{eq:5_6}
\end{equation}
Both bounds are not very restrictive. We discuss next the scattering of neutrinos on electrons or nuclei which are sensitive to a larger range of the coupling constant.

\section{Elastic neutrino--electron scattering\label{sec6}}

A very clear channel to search for effects of the new terms are neutrino--electron and antineutrino--electron scattering. The model is well defined and the signature for the effects are recoiling electrons in the forward direction. The vertices are determined by the simple interaction in~\eqref{eq:4_4}. The amplitude originates from couplings of $X_\mu$ to the standard neutrinos in the beam

\begin{eqnarray}
\mathcal{M}_e=i\left(\frac{g_Xc_\alpha}{4}\right)^2Y_\varphi^2\frac{2(1-2s_W^2)}{m_X^2+2T_em_e}\bar u_\nu\gamma^\mu(1-\gamma^5) u_\nu\bar u_e\gamma_\mu u_e.\label{eq:6_1}
\end{eqnarray}
This amplitude interferes with the amplitude from the standard model. For reasons that we explain below, we assume that the new contribution to the cross section is at most $20\%$, which determines the new coupling as a function of $m_X$. 

\begin {table}
\begin{center}
\begin{tabular}{ |c|c| }
  \hline
   $m_X$ in MeV & $\sqrt{\lambda}=\dfrac{g_Xc_\alpha}{4}Y_\varphi$ \\\hline
  $1$ & $1.5\times10^{-6}$ \\
	$5$ & $4.8\times10^{-6}$ \\
  $10$ & $0.9\times10^{-5}$ \\
  $100$ & $0.9\times10^{-4}$ \\
  1000 & $0.9\times10^{-3}$ \\
  \hline
\end{tabular}
\caption {Upper bound of the effective coupling for elastic antineutrino--electron scattering as a function of the light mediator mass.}\label{tab_r}
\end{center}
\end {table}

Two experiments CHARM~\cite{Vilain:1993kd} and TEXONO~\cite{Deniz:2009mu} measured the leptonic reactions at two different energies. The CHARM Collaboration~\cite{Vilain:1993kd} reported the differential cross section as a function of the recoiling energy and showed that the experimental points follow the distribution of the standard model. The error bars, however, are larger than $25\%$ and the normalization is in arbitrary units. The TEXONO Collaboration~\cite{Deniz:2009mu} obtained results at lower energies $3<E_\nu<8$ MeV and reported results integrated over the recoiling energy of the electrons. The error bars for the integrated cross section are again $25\%$ or larger.

The MINERvA Collaboration collected a significant number elastic scatterings of neutrinos and antineutrinos on electrons. For the various flavors in the beam they used standard model cross sections  in order to constraint the low~\cite{Park:2015eqa} and medium energy~\cite{minerva} flux. Including the additional contribution, like the one in this article, will slightly modify the flux. Finally, efforts for improving the results of the COHERENT experiment led to a new measurement of the quenching factor (QF) for CsI(Na) and an assessment of previous calibrations~\cite{Collar:2019ihs}. This motivated new theoretical analysis of the COHERENT data~\cite{Papoulias:2019txv,Khan:2019cvi,Giunti:2019xpr}, which brought the central value of the Weinberg angle close to the world average with a combined error of 20\% ($\sin^2\theta_W=0.238\pm0.045$). Under these circumstances it is justified to adopt the approach that an additional contribution to the cross section of $20\%$ is not excluded.

In the model the additional contribution modifies the vector coupling at the electron vertex $g_V\rightarrow g_V+\kappa$ with

\begin{equation}
\kappa=\frac{\sqrt{2}}{G}\left(\frac{g_Xc_\alpha}{4}\right)^2Y_\varphi^2\frac{1}{m_X^2+2m_eE_\nu x}=\frac{\sqrt{2}}{G}\frac{\lambda}{m_X^2+s x}
\end{equation}
with $\lambda=(g_Xc_\alpha/4)^2Y_\varphi^2$, $x=T_e/E_\nu$ and we set $2(1-2s_W^2)=1$.

\begin{figure}
\centering
\includegraphics[width=0.7\textwidth]{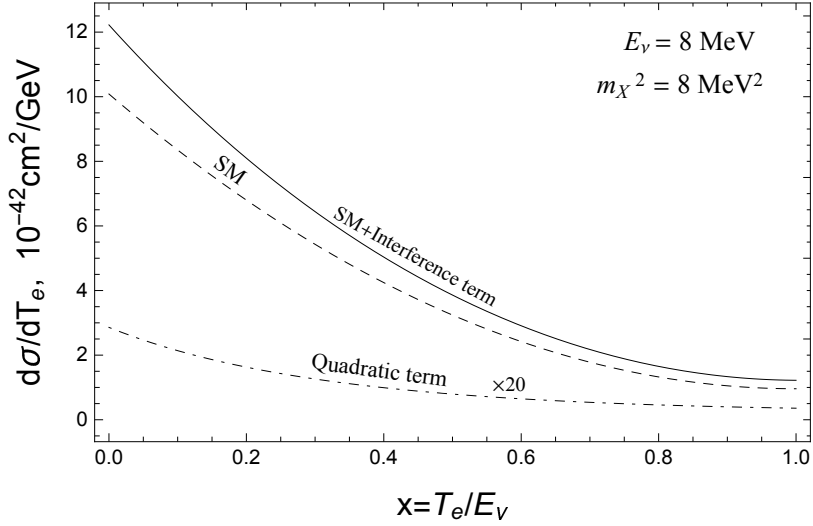}
\caption{Differential cross section for elastic antineutrino--electron scattering. The dashed curve is for the standard model. The solid curve is the standard model plus interference term. The dot-dashed curve is 20 times the quadratic term. The $X_\mu$ boson mass squared is taken to be 8~MeV$^2$, the incoming neutrino energy is 8 MeV.}
\label{fig:15}
\end{figure} 

Next, we address the question of determining the coupling constant. Since the TEXONO experiment integrated over the the electron spectrum we shall assume that the new terms integrated over the electron recoiling energy are smaller than $20\%$ of the standard model contribution. This gives the condition

\begin{equation}
\frac{\dfrac{\sqrt{2}}{G}\lambda\int_0^{1}[(g_V-g_A)+(g_V+g_A)(1-x)^2]\dfrac{1}{m_X^2+sx}dx}{(g_V-g_A)^2+\frac{1}{3}(g_V+g_A)^2}\leq0.1.
\end{equation}
In this relation we kept only the interference term and integrated over $x$. For $m_X^2\leq s$ we use the exact integral and for $m_X^2\gg s$ we approximate the propagator by $1/m_X^2$. A consequence of the condition are the values for the coupling constant as a function of $m_X$ presented in Table~\ref{tab_r}. They indicate the sensitivity that can be reached by experiments. With the couplings in Table~\ref{tab_r} the cross section for elastic scattering is determined including the contribution of the new terms. The interference introduces a new term to the differential cross section

\begin{equation}
\left(\frac{d\sigma}{d T_e}\right)_{\text{INT}}=\frac{\sqrt{2}Gm_e}{\pi}\lambda[(g_V-g_A)+(g_V+g_A)(1-x)^2]\frac{1}{m_X^2+sx}.
\end{equation}
The dependence on the recoiling energy including the propagator of a light $X_\mu$ boson determines the shape of the curve. In Fig.~\ref{fig:15} we chose an energy $E_\nu=8$ MeV and a mass $m^2_X=8$~MeV$^2$. The standard model cross section is shown with a dashed curve. We assume the interference to be constructive and their sum is the solid curve. There is a small change in the shape of the spectrum with the additional cross section being sizable at~$x\leq0.5$. In the same figure we plotted the quadratic addition to the cross section but this time multiplied by a factor of twenty in order to be visible. This contribution is flatter but smaller. The same curve also gives the recoiling spectrum for reaction~\eqref{eq:1in} when Majorana neutrinos are present in a beam.

\section{Coherent elastic neutrino--nucleus scattering\label{sec7}}
In the extended model there are  also new couplings of the quarks; the couplings for the up and down quarks have the following from:

\begin{eqnarray}
\mathcal{L'}(X)=X_\mu\frac{g_Xc_\alpha}{4}Y_\varphi\left\{2c_W^2\left(\bar u\gamma^\mu u-\bar d\gamma^\mu d\right)-\frac{2}{3}s_W^2\left(\bar u\gamma^\mu u+\bar d\gamma^\mu d\right)\right\}. \label{eq:32n}
\end{eqnarray}
The dominant term for coherent scattering of neutrinos on nuclei is the one proportional to the isovector current. The additional amplitude is

\begin{equation}
\mathcal{M}_C=i\frac{\mu}{m_X^2+2MT}\bar u_\nu\gamma_{\mu}(1-\gamma^5)u_\nu\left[\bar u\gamma^\mu u+\bar d\gamma^\mu d\right]\label{eq:34n}
\end{equation}
with $\mu=(g_Xc_\alpha)^2Y_\varphi^2s_W^2/24$ and  the quark vertex being the baryon current, whose matrix element between nuclei gives the atomic number times a from factor.

The new amplitude contributes coherently to the cross section producing

\begin{equation}
\frac{d\sigma}{dT}=\left[1+\frac{2\sqrt{2}}{G}\frac{A}{Q_W}\frac{\mu}{m_X^2+2MT}\right]^2\left(\frac{d\sigma}{dT}\right)_{\text{SM}}. \label{eq:33n}
\end{equation}
Here $(d\sigma/dT)_{\text{SM}}$ is the standard model cross section with  the recoiling kinetic energy of the nucleus~$T$, $Q_W=[(1-4s_W^2)Z-N]$ and $A$ is the atomic number of the nucleus. Allowing $1\sigma$ uncertainty for the Weinberg angle brings a change on the value of the cross section larger than $20\%$. Thus assuming the contribution of the new amplitude to be smaller than $10\%$ of the amplitude for the standard model restricts the coupling constant to be

\begin{equation}
4\sqrt{\mu}<10^{-4}\,\,\,\,\,\,\text{for}\,\,\,\,\,\,m_X^2\ll2MT.\label{q:37n}
\end{equation}
For the COHERENT experiment~\cite{Akimov:2017ade,Scholberg:2015} the energy range is $16<E_\nu<53$ MeV and the values for~$2MT$ correspond to the higher values in Table~\ref{tab_r}. In addition, the model has only one hypercharge and thus the bound in~\eqref{q:37n} is consistent with the values of the table.

\begin{figure}
\centering
\includegraphics[width=0.7\textwidth]{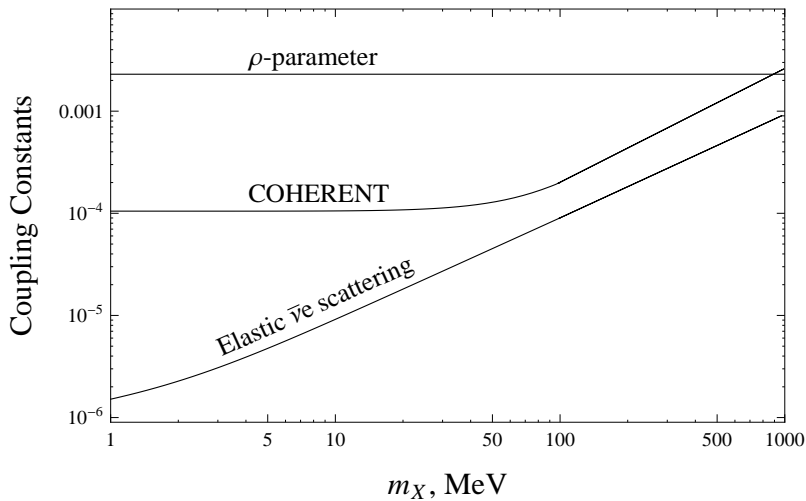}
\caption{Upper limits for the coupling constants as functions of $m_X$ obtained from various experiments. The upper curve is from the $\rho$-parameter; middle curve from COHERENT~\cite{Akimov:2017ade} and lower curve from elastic antineutrino--electron scattering~\cite{Deniz:2009mu}. The curves for $\bar\nu e^-$ scattering and coherent scattering show the limits for the couplings $\sqrt{\lambda}$ and $2\sqrt{\mu}$, respectively.}
\label{fig:26}
\end{figure}

In Fig.~\ref{fig:26} we plotted the exclusion regions for the coupling constants from various experiments. The upper curve corresponds to the bound from~\eqref{eq:restr_c} and is rather weak. The middle curve is from coherent elastic neutrino--nucleus scattering. For values of $m_X^2\lesssim2MT\approx 3.6\times10^{3}$~MeV$^2$ the curve is flat and the mass of $X_\mu$ becomes visible for $m_X\geq 60$~MeV. The most restrictive values come from $\bar\nu_e+e^-\rightarrow\bar\nu_e+e^-$ scattering where the momentum transfer squared for the electrons $2m_eT_e$ is very small of $\mathcal{O}(8~\text{eV}^2)$ and the excluded region for the coupling begins to grow at a small value of $m_X$. Non-standard neutrino couplings to a light boson $X$ have been studied and bounds were obtained for phenomenological couplings~\cite{Lindner:2016wff,Abdullah:2018ykz,Billard:2018jnl}. They are comparable to our bounds in Fig.~\ref{fig:26}.
Any differences that may exist between the articles can be accounted for by the different approaches (universal couplings versus gauge couplings) and methods for analyzing the data.

In our discussion we assumed constructive interference. In Fig.~12 of~\cite{Billard:2018jnl} the relative signs of the standard model and the new contribution are opposite producing a significant destructive interference. In Fig.~14 of the same article the excluded region of couplings for $\nu-e$ elastic scattering is close to the values in our Table~\ref{tab_r}.

\section{Summary and conclusions\label{conclusion}}
We developed a model with right-handed neutrinos being Majorana particles coupled to a neutral gauge boson. The coupling is pure axial-vector. This takes place in a multiplicative group with its content being invariant under chiral gauge transformations. The new boson, $X$, mixes weakly with particles of the standard model, which makes the detection of new effects difficult.

Masses for the particles are introduced with the standard Higgs doublet and a new scalar singlet. The mixings of the gauge bosons have common properties with earlier studies of $U(1)_X$ models~\cite{Fayet:1989mq,Appelquist:2002mw,Fayet:2016nyc,Lindner:2018kjo}. The consistency of the model relates the hypercharges of quarks and leptons and reduces them to one independent charge; thus several reactions are correlated.

Bounds on the couplings from the $\rho$-parameter and decays of the $Z$ boson to invisible $N_M$ particles are presented but are not the most restrictive. For this reason we reviewed the experimental results for (anti)neutrino--electron scattering and concluded that unknown contributions as large as $20\%$ are still possible.
 Then we included the additional terms to the $g_V$ couplings of the electron and plotted the changes in the cross section. The detection of new effects requires an independent determination of neutrino or antineutrino flux in addition to the precise measurement of the recoiling energy of electrons.
   
	A second process is coherent scattering of neutrinos on atomic nuclei. The dependence of the cross section on the small recoiling energy will be the same as in the standard model. The corresponding amplitude has a pure baryonic current coupled to quarks which enhances the cross section. This is an advantage of the model, that the couplings are related through gauge invariance and correlations among various reactions are possible. For coherent scattering the searches for new effects must rely on the absolute rate of the cross section.
  
	The observable effects we discussed for the reactions rely mostly on the mixing of $X$ with the $Z$ boson, which produces couplings to standard neutrinos. To observe reactions induced by the Majorana neutrinos it is necessary to create them in operating beams. This may occur in beam dump experiments~\cite{Alekhin:2015byh}, where reactions with the $X_\mu$ in intermediate states can create them. Spallation neutron sources have the advantage of producing new particles. A second possibility is provided by the mass matrix where Majorana states are created through oscillations. Recent results from the MINOS+ Collaboration reported results~\cite{Adamson:2017uda} that eliminate this possibility. MINOS+ is a long base line experiment with two detectors; a near detector 1.04~km from the NUMI target and a far away detector at 735~km. The detectors were designed to search for disappearance and set a limit $\sin^2{2\theta_{24}} < 0.008$ and $\Delta m^2_{41}>10^{-2}$~eV$^2$. The probability for generating an invisible component in the beam is very small. 

Finally, the Majorana neutrinos with low mass will decouple in the early universe and at the same time will satisfy the Lee--Weinberg bound~\cite{Lee:1977ua}. An example for $m_X=10$~MeV and a light Majorana is discussed on page~233 of~\cite{Boehm:2003hm}. The value for the product of coupling required in order to produce the relic abundance of dark matter is very close to the bound for the coupling constant of the TEXONO experiment plotted in Fig.~\ref{fig:2}.

\acknowledgments
We wish to thank Drs. F.~C.~Correia, P.~Fayet and W.~Rodejohann for valuable comments they made on the earlier version of the article. One of us (I.A.) was partly supported by the Program of fundamental scientific research of the Presidium of the Russian Academy of Sciences "Physics of fundamental interactions and nuclear technologies".


\end{document}